\documentclass[sigconf,nonacm]{acmart}

\usepackage{graphicx}
\usepackage{subcaption}
\usepackage{siunitx}

\AtBeginDocument{%
  }

\renewcommand\footnotetextcopyrightpermission[1]{}
\acmDOI{}
\acmConference[KDD '25]{Make sure to enter the correct
  conference title from your rights confirmation email}{Aug 03--07,
  2025}{ON, Toronto}
\acmISBN{}

\begin{document}

\title{Trading with the Devil: Risk and Return in Foundation Model Strategies}


\author{Jinrui Zhang}
\affiliation{%
  \institution{Tsinghua University}
  \city{Beijing}
  \country{China}}
\email{zhangjr23@mails.tsinghua.edu.cn}








\begin{abstract}
Foundation models—already transformative in domains such as natural language processing—are now starting to emerge for time‐series tasks in finance. While these pretrained architectures promise versatile predictive signals, little is known about how they shape the risk profiles of the trading strategies built atop them, leaving practitioners reluctant to commit serious capital. In this paper, we propose an extension to the Capital Asset Pricing Model (CAPM) that disentangles the systematic risk introduced by a shared foundation model—potentially capable of generating alpha if the underlying model is genuinely predictive—from the idiosyncratic risk attributable to custom fine‐tuning, which typically accrues no systematic premium. To enable a practical estimation of these separate risks, we align this decomposition with the concepts of uncertainty disentanglement, casting systematic risk as epistemic uncertainty (rooted in the pretrained model) and idiosyncratic risk as aleatory uncertainty (introduced during custom adaptations). Under Aleatory Collapse Assumption, we illustrate how Monte Carlo dropout—among other methods in the uncertainty‐quantization toolkit—can directly measure the epistemic risk, thereby mapping trading strategies to a more transparent risk–return plane. Our experiments show that isolating these distinct risk factors yields deeper insights into the performance limits of foundation‐model‐based strategies, their model degradation over time, and potential avenues for targeted refinements. Taken together, our results highlight both the promise and the pitfalls of deploying large pretrained models in competitive financial markets.
\end{abstract}




\settopmatter{printacmref=false}
\maketitle

\section{Introduction}
Recent advances in machine learning has transformed the landscape of modern financial markets \cite{hajj2023unveiling, cao2023finance}, with applications spanning single asset timing \cite{nelson2017stock, selvin2017stock}, portfolio optimization \cite{zhang2020deep}, and even market making \cite{zhong2020datadriven}. In parallel, trading models have evolved from single‐variate transformers \cite{malibari2021predicting, ding2020hierarchical}, reinforcement learning agents \cite{deng2017deep}, to complex multi‐modal systems with LLM agents \cite{sun2023mastering}. While these approaches have demonstrated promise in capturing complex market signals, risk management remains elusive in these blackbox strategies. Some tentative solutions try to incorporate uncertainty information into the model itself \cite{chauhan2020uncertaintyaware, sun2022quantitative}. As Goodhart’s Law \cite{goodhart1984problems} reminds us, embedding risk measures within the same predictive model can inadvertently inflate systemic vulnerabilities \cite{boucher2014risk, danielsson2016model}—\textit{When a measure becomes a target, it ceases to be a good measure}.

Meanwhile, large pretrained models are quickly gaining traction in finance \cite{wu2023bloomberggpt, Yang2023FinGPTOF}. Newly emerging large time‐series models (LTSM) similarly undergo fine‐tuning on specific market data to bootstrap predictive performance for tasks like mid‐price prediction or intraday signals \cite{fu2024financial}. Yet the widespread adoption of foundation models begs the question: \textit{do multiple trading strategies—fine‐tuned from the same underlying pretrained network—exhibit correlated performance and systemic risk?}

In classical finance, the Capital Asset Pricing Model (CAPM) decomposes a portfolio’s total risk into systematic (market‐driven) and idiosyncratic (asset‐specific) components \cite{sharpe1964capital, fama2004capital}. Interestingly, uncertainty disentanglement in machine learning splits a model’s predictive uncertainty into epistemic (model‐related) and aleatory (data‐intrinsic) parts \cite{der2009aleatory}. This symmetry motivates a fundamental question: \textit{Can we reconcile these two viewpoints so that each trading strategy’s foundation‐model‐driven risk lines up with CAPM’s notion of systematic exposure, and each fine‐tuning quirk aligns with idiosyncratic or aleatory uncertainty?} If so, we could leverage established financial risk theories to deepen our understanding of how large pretrained models collectively shape market dynamics.

In this paper, we bridge the lens of CAPM‐style risk analysis with uncertainty disentanglement to provide a cohesive framework for analyzing trading strategies fine‐tuned from large pretrained (foundation) models. Specifically:

\begin{figure*}
    \centering
    \includegraphics[width=\textwidth]{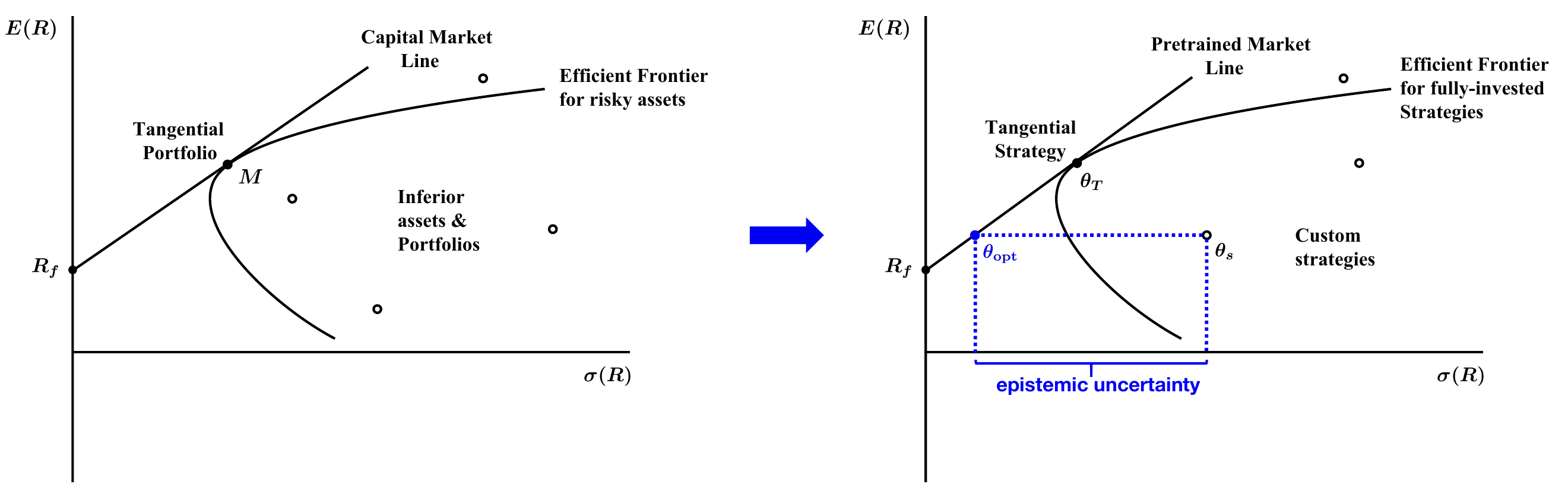}
    \caption{A High‐Level View: From the Classical CAPM to the “Foundation‐Model” CAPM. On the left, we show the familiar Capital Market Line (CML) linking the risk‐free asset \(R_f\) to a tangential portfolio \(M\), which represents the fully diversified market portfolio in CAPM theory. On the right, we illustrate our foundation‐model extension, where the tangential portfolio \(M\) is replaced by the tangential strategy \(\theta_T\) that fully exploits the pretrained network. Custom fine‐tuned strategies \(\theta_s\) may deviate from the Pretrained Market Line (PML) if they bear additional, unpriced (idiosyncratic) variance. We posit a hypothetical optimal strategy \(\theta_{\mathrm{opt}}\), which shares the same expected return but carries only epistemic risk.  The distance in standard deviation between \(\theta_s\) and \(\theta_{\mathrm{opt}}\) captures the extra variance that could be avoided by more efficient adaptation of the foundation model. This figure foreshadows our framework for adapting CAPM principles to foundation‐model trading, forming the conceptual basis for Sections 3.1 and 3.2.}
    \Description{Side-by-side risk–return plots compare the classical and foundation-model CAPM approaches. The left chart shows the risk-free asset Rf, the tangential portfolio M on the Capital Market Line, and some inferior assets below it. On the right, the tangential portfolio is replaced by a tangential strategy from a Pretrained Market Line. Custom strategies deviate from that line and carry extra idiosyncratic variance. A hypothetical optimal strategy theta_opt shares the same expected return but has only epistemic risk, illustrated by the horizontal distance between theta_s and theta_opt.}
\end{figure*}
    
\textbf{A CAPM‐Inspired Framework for Foundation Model Trading.} Our framework starts by defining the Pretrained Market Line (PML) in mean-variance space, which represents the best risk-return trade-off investors can achieve leveraging foundation model’s predictive edge. By analogy with the conventional Capital Market Line (CML), which represents the tangency portfolio combined with the risk-free asset, the PML emerges as the tangent line connecting the risk-free rate to the efficient frontier of fully-invested strategies. This frontier is shaped by the optimally-tuned pretrained model’s ability to refine return forecasts and covariance estimates, thereby shifting the tangency portfolio to a higher Sharpe ratio regime. We demonstrate that if the foundation model embodies informative alpha, its signals enable a steeper PML (higer Sharpe ratio) than the classical CML, as the model-driven tangency portfolio dominates the market portfolio in mean-variance space. This results in a superior risk-return trade-off, where investors can achieve higher expected returns per unit of risk by leveraging the foundation model’s predictive edge.

\textbf{Uncertainty‐Based PML Estimation.} Although the PML offers an elegant conceptual benchmark, its direct empirical estimation is non‐trivial—no less difficult than identifying the fully-diversified market portfolio in standard CAPM. We therefore devise a Bayesian uncertainty‐disentanglement method to approximate the PML for real‐world strategies. Under a key assumption—namely, that any given fine‐tuned strategy exhibiting certain risk and return can be matched by an ideal strategy with identical return yet only the foundation model’s epistemic (shared) risk component—we estimate how much of a strategy’s variance is non‐diversifiable and thus priced on the PML. This approach leverages modern uncertainty quantization (e.g., Monte Carlo dropout) to differentiate between epistemic risk that arises from the pretrained model itself and aleatory risk introduced by suboptimal or idiosyncratic customizations.
    
\textbf{Empirical Validation.} Finally, we test our framework on popular pretrained large time series across multiple asset classes, including US equities and cryptocurrencies. We estimate the PML in a rolling‐window fashion, offering insights into how alpha evolves over time as market conditions shift. 
    
Taken as a whole, our framework paves the way for a more transparent evaluation of foundation‐model‐based trading. Rather than focusing exclusively on alpha generation, we highlight how risk—particularly systemic risk shared by many market participants using the same base model—can propagate through financial markets and, in line with CAPM principles, help explain the returns observed. In the sections that follow, we detail our CAPM‐inspired formulation, present an uncertainty‐based risk‐measurement strategy (with Monte Carlo dropout as one concrete instantiation), and validate the approach empirically on multiple large time‐series architectures. We conclude by discussing the limitations of our work and outlining promising directions for future research—specifically, extending our empirical tests to multivariate and portfolio‐level strategies, addressing the latency–scaling trade‐offs inherent in larger models, and cross-model analysis of shared risk factors—to guide the safe scaling of foundation models in finance.

\section{Background}

Our work builds upon two primary foundations: the classical Capital Asset Pricing Model (CAPM), grounded in mean–variance optimization, and model uncertainty disentanglement, grounded in bayesian methodologies. which separates distinct sources of predictive risk. We summarize each in turn.

\subsection{Mean–Variance Space and the Capital Market Line} 

The origins of CAPM lie in Markowitz’s mean–variance optimization \cite{markowitz1952portfolio}. In this framework, investors select portfolios by balancing expected returns against variances of returns. Specifically, if $\mathbf{w}$ is the vector of portfolio weights on n risky assets (with a covariance matrix $\Sigma$ and expected return vector $\boldsymbol{\mu}$), then an investor typically solves:
\begin{equation}
\min_{\mathbf{w}} \ \mathbf{w}^\top \Sigma \,\mathbf{w}
\quad 
\text{subject to} 
\quad
\mathbf{w}^\top \boldsymbol{\mu} = \mu_{\mathrm{target}}, 
\quad
\mathbf{1}^\top \mathbf{w} = 1
\end{equation}
where $\mu_{\mathrm{target}}$ is the desired portfolio return and $\mathbf{1}$ is the all‐ones vector (for fully invested portfolios). Tracing out $\mu_{\mathrm{target}}$ over all feasible values produces the efficient frontier of portfolios with minimal variance for a given expected return.

\textbf{The Capital Market Line (CML)}: When a risk‐free asset with return $r_f$ is introduced into the opportunity set, the efficient frontier is reduced to a single tangential portfolio $M$ that maximizes the Sharpe ratio. All optimal portfolios then lie on the Capital Market Line (CML), described by:
\begin{equation}
E[r_p] 
\;=\; 
r_f 
\;+\; 
\Bigl(\tfrac{E[r_M]-r_f}{\sigma_M}\Bigr)\,\sigma_p
\end{equation}
where 
$E[r_p]$ is the expected return of the portfolio,
$r_f$ is the risk‐free rate,
$E[r_M]$ is the expected return of the tangential (market) portfolio $M$
$\sigma_M$ is the standard deviation of $M$, and
$\sigma_p$ is the standard deviation of the chosen portfolio $p$.

All portfolios on the CML can be seen as combinations of the risk‐free asset and the tangential portfolio $M$. The slope of CML $({E[r_M] - r_f}) /\ {\sigma_M}$ is referred to as the market Sharpe ratio.

\subsection{Classic CAPM Formulation}

Building on mean–variance optimization, the Capital Asset Pricing Model (CAPM) \cite{sharpe1964capital} posits that any asset $i$ (or portfolio $p$) has an expected return determined by its systematic exposure $\beta_i$ to the market:
\begin{equation}
E[r_i]
\;=\; 
r_f 
\;+\; 
\beta_i 
\bigl(E[r_m] - r_f\bigr),
\qquad
\beta_i 
\;=\;
\frac{\mathrm{Cov}(r_i,\, r_m)}{\mathrm{Var}(r_m)}
\end{equation}
Under this view, the market is identified with the tangential portfolio $M$. The associated variance decomposition for asset (or portfolio) $i$ is:
\begin{equation}
\sigma_i^2
\;=\;
\beta_i^{2}\,\sigma_m^2
\;+\;
\sigma_{\varepsilon}^2    
\end{equation}
where $\beta_i^{2}$, $\sigma_m^2$ is the systematic (market) risk and $\sigma_{\varepsilon}^2$ is the idiosyncratic (diversifiable) risk. The classical CAPM thus separates total variance into two parts—only the systematic portion merits a risk premium in equilibrium.

\subsection{Bayesian Perspective and Model Uncertainty}

Beyond the realm of finance, Bayesian methodologies have proven instrumental for analyzing predictive uncertainty in machine learning, offering tools to separate and quantify distinct uncertainty sources. From a Bayesian perspective, a model’s predictive distribution,
\begin{equation}
    p(y \mid x,\mathcal{D}) \;=\; \int_{\Omega} p(y \mid x,\Theta)\,p(\Theta \mid \mathcal{D}) \,d\Theta
\end{equation}
encodes both the likely values of the output $y$ and the uncertainty surrounding those values, given an input $x$ and training data $\mathcal{D}$. Although this integral succinctly characterizes predictive risk, it remains analytically intractable in most neural network settings. 

A common strategy for approximating this predictive distribution is provided by \textit{Monte Carlo (MC) Dropout} \cite{gal2016dropout}. Originally introduced as a regularization scheme, dropout randomly “switches off” neurons during training. In MC-Dropout, this randomness is retained at inference time, thereby sampling different model configurations from an approximate posterior $\Theta_t \sim q(\Theta \mid \mathcal{D})$). By aggregating predictions across multiple forward passes, one obtains not only a mean prediction but also an empirical variance that reflects model’s uncertainty. 

In the field of uncertainty disentanglement, researcher further distinguishes model uncertainty between two principal sources\cite{Hora1996AleatoryAE}:

\begin{itemize}
    \item \textbf{Epistemic Uncertainty (model‐based)}: Emanating from limited model knowledge or insufficient training data, epistemic risk can be mitigated through additional information or improved modeling. In time-series trading, this might manifest as sub-optimal fine tuning and lack of finnancial specific priors. This type of uncertainty reflects how sensitively the network’s predictions depend on its parameters: more pronounced variability across forward passes indicates that the model’s beliefs about $f(x)$ are unstable or underdetermined by the available data.

    \item \textbf{Aleatory Uncertainty (data‐intrinsic)}: Stemming from inherent noise or stochasticity in the data‐generating process, aleatory risk cannot be reduced by collecting more data or refining the model. In time‐series trading, this might manifest as unpredictable shocks or volatility spikes that no model—however sophisticated—could reliably foresee.
    
\end{itemize}

By applying such Bayesian-inspired techniques, one can more clearly distinguish between variance due to fundamental randomness (aleatory) and that stemming from incomplete model knowledge (epistemic). As we shall demonstrate in subsequent sections, recognizing this distinction is essential for CAPM-style analyses of systematic versus idiosyncratic risk in foundation-model-based trading.

\section{CAPM for Foundation Model Trading}
\subsection{Notation and Strategy Instances}
We begin by formalizing the basic objects of our framework. At a high level, 
a pretrained backbone $\theta$ induces a family of fine-tuned strategies through 
different knobs of adaptation and execution. Each such strategy produces a 
return distribution characterized by mean and volatility.

\begin{definition}[Backbone family]
Given a pretrained backbone $\theta$, define the strategy family 
\[
\mathcal{S}(\theta) = \{\,(\theta,\phi,\kappa): \ \phi\in\Phi,\ \kappa\in\mathcal{K}\,\},
\]
where $\phi$ denotes fine-tuning controls (e.g., data subsets, loss weights, 
regularization, LR schedule) and $\kappa$ execution controls (e.g., stop-loss, 
take-profit, sizing). Each $\theta_s\in\mathcal{S}(\theta)$ produces a return time series $\{r_{s,t}\}_{t=1}^T$ with mean $\mu_s$ and stdev $\sigma_s$. We denote cost\,adjusted returns by $\tilde r_{s,t}$ after fees/slippage.
\end{definition}

\subsection{Mean-variance Equilibrium : from CML to PML}

Classical Markowitz theory yields a mean–variance efficient frontier; adding a risk‐free asset and allowing risk‐free borrowing/lending produces the \emph{Capital Market Line (CML)}, the straight line through $(\sigma{=}0,\,E[r]{=}r_f)$ and the unique tangency portfolio $T_f$. Tobin’s separation theorem implies any optimal choice is a mixture of $r_f$ and $T_f$. In the standard CAPM setting, $T_f$ is identified with 
the market portfolio $M$.

In foundation model analogue, the role of $M$ is played by an ideal fine-tuned 
strategy $\theta_T$ that best exploits the predictive edge from the backbone model $\theta$. 
We now formalize this tangential strategy and its induced efficient frontier.

In foundation model trading, suppose one has a universe of potential trading signals, all generated or informed by a single pretrained network $\theta$. Each fine‐tuned strategy $\theta_i$ can selectively engage or stays out of the market (e.g., by thresholding borderline signals), thereby reserving some portion of capital in the risk‐free asset $R_f$. The resulting efficient set of foundation‐based strategies (PML) thus originates at $R_f$ and passing through a tangential strategy $T_f$. In our context, $T_f$ is identified with the fine-tuned strategy that most effectively exploit the foundation model’s signals and always fully-invested in risky assets ($\mathbf{1}^\top \mathbf{w} = 1$), denoted $\theta_{\mathrm{T}}$.

\begin{definition}[Tangential backbone strategy]
Let $\theta_T\in\mathcal{S}(\theta)$ be the \emph{fully invested} 
(no risk-free mixing) strategy maximizing the Sharpe ratio:
\[
\mathrm{SR}_\theta \;=\; \frac{E[\tilde r_{\theta_T}] - r_f}{\operatorname{sd}(\tilde r_{\theta_T})}.
\]
\end{definition}

\textbf{Complete Agreement for Equilibrium}: Among the many simplifying assumptions in CAPM, the idea of "complete agreement"—where all investors observe the same distribution of asset returns and select mean–variance-efficient portfolios—finds an interesting parallel in the context of foundation-model trading. We note that multiple practitioners are obliged to fine-tune on the same pretrained weights, given that retraining these models from scratch is prohibitively expensive. Consequently, these fine‐tuned strategies, despite their superficial differences, share largely similar predictive signals, thereby forming an approximate “agreement” about the future markets. In other words, it is not that practitioners independently converge on identical beliefs, but rather that the hefty foundation model itself imposes a common informational baseline.

\textbf{PML and Optimal Sharpe Ratio}: Drawing analogy from CML,  Now we formally identify the optimal sharpe ratio of a pretrained model, which is also the slope of PML:
\begin{equation}
\text{SR}_\theta \;=\; \frac{ E\bigl[r_{\theta_T}\bigr] \;-\; r_f }{ \sigma\bigl(r_{\theta_T}\bigr) }
\end{equation}

The sharpe ratio $\text{SR}(\theta)$ capturing how much excess return $\theta_T$ delivers per unit of risk. Each optimal foundation‐based strategy that mixes some proportion of $R_f$ with $\theta_T$ will reside on this PML, achieving an expected return and standard deviation consistent with standard mean–variance theory:
\begin{equation}
E[r_\mathrm{opt}] \;=\; r_f \;+\; \sigma_\mathrm{opt} \times \text{SR}_\theta    
\end{equation}

\textbf{The Challenge of Identifying $\theta_T$}:  Much as the CAPM struggles with the market proxy problem \cite{sharpe1964capital}—the “true” market portfolio is theoretically elusive—our framework also faces a “foundation tangential” proxy challenge. While $\theta_T$ is posited to be the unique all‐risky, mean–variance‐efficient strategy of signals based on the pretrained model, in practice, one rarely has perfect knowledge of which exact combination of signals, fine-tuning technique, and hyperparameters yields the “true” $\theta_T$. Consequently, empirical tests must rely on approximations or proxies for $\theta_T$, analogous to how empirical finance employs broad market indices as stand‐ins for the market portfolio. From a financial standpoint, one would use an information-leaked model  $\theta_\mathrm{leak}$, which peeks at future data.  Such a leaked model can serve as a simplistic upper-limit construction: by capitalizing on prior knowledge of upcoming market conditions. In the next section, we will show how uncertainty disentanglement can help pinpoint a better estimation on $\theta_T$’s risk-return performance.

\subsection{PML Estimation via Uncertainty Disentanglement}

In the preceding section, we introduced the notion of the Pretrained Market Line (PML) and posited a tangential fine‐tuned strategy, $\theta_T$, that fully exploits the foundation model’s predictive capabilities. Now, we address the practical challenge of estimating the PML using observed suboptimal strategies.

Let $\theta_s$ denote the parameters of a custom fine‐tuned model, producing trading signals that yield (random) returns $r_s$. We characterize $r_s$ by its expected return $E[r_s]$ and standard deviation $\sigma_s$. To estimate how much of $\sigma_s$ is genuinely systematic (i.e., priced on the PML), we adopt the following assumption:

\textbf{Assumption 3.1.} \textit{\textbf{Aleatory Collapse}. In a competitive market, any readily observable aleatory uncertainty is swiftly arbitraged away, due to its' correlation with return. As a result, an uncertainty estimator rapidly converges to an epistemic uncertainty estimator.} 

In other word, any custom strategy $\theta_s$ can be matched by a hypothetical optimal strategy, $\theta_s^\mathrm{opt}$, that attains the return but exposes only the portion of $\sigma_s^2$ that is unavoidable (i.e., the aleatory or systematic risk tied to the foundation model). All additional variance is deemed idiosyncratic and can be “diversified away” by better exploitation of the pretrained signals.

With this assumption, we use a Bayesian risk‐quantization approach—specifically, MC Dropout—to disentangle $\sigma_s$. Concretely, given fine-tuned strategy $\theta_s$, we first obtain $K$ dropout activated variants $\{ \theta^{(k)}_s \}^K_{k=1}$. Denoting the backtested returns as $\{ r^{(k)}_s \}^K_{k=1}$, the cross‐sample variance $\sigma_\mathrm{MC}^2$ approximates the epistemic component:
\begin{equation}
\mu_{\text{MC}} = \frac{1}{K} \sum_{k=1}^K \hat r^{(k)}_s, \quad 
\sigma^2_{\text{MC}} = \frac{1}{K} \sum_{k=1}^K \left( \hat{r}_s^{(k)} - \mu_{\text{MC}} \right)^2     
\end{equation}

\textbf{Regression-Based Estimation of PML}:  Given multiple custom strategies $\{\theta_{s_1}, \theta_{s_2}, \dots\}$, we can compile an empirical set of $\{\,\sigma_{s_i},\,E[r_{s_i}]\}$ on the risk-return plane. To estimate the underlying slope of the PML (Foundation Sharpe Ratio), we start by subtract out the idiosyncratic portion from  using the bayesian decomposition above, leaving a “priced” standard deviation $\{\,\sqrt{\sigma^2_{s_i} - \sigma^2_\mathrm{epi}},\,E[r_{s_i}]\}$. Next, perform a linear fit relating $E[r_{s_i}]$ to its risk. The resulting slope is an estimate of $\text{SR}_{\theta}$, the maximum Sharpe ratio that the foundation model could theoretically provide, absent extraneous (idiosyncratic) risk. Thus, by analyzing many fine‐tuned strategies in the risk-return plane—once purged of their unpriced variance—we glean an empirical vantage point on the pretrained market line.

\textbf{Foundation Model Decay}: In practice, this bounding procedure can be recalculated on a rolling basis, furnishing a powerful mechanism to dissect both trading crowding effects and the potential decay of foundation models. One may observe the upper limit on $\mathrm{SR}_\theta$ gradually drifting downward—signaling the foundation’s waning alignment with evolving market conditions—and simultaneously witness the gap between optimal $\mathrm{SR}_\theta$ and custom $\mathrm{SR}_\mathrm{s}$ widening as the strategy’s signals become more broadly disseminated. Sharp declines in performance may also emerge when new, more potent foundation models appear. We believe this dynamic framework, illustrated in \autoref{fig:decay}, opens fertile ground for deeper exploration of ephemeral alpha, model obsolescence, and market equilibria.

\textbf{Connection with Ensembling}: Our pricing model also naturally connects with ensemble‐based strategies, where multiple models—or experts—are combined to enhance predictive accuracy \cite{yang2020deep, nti2020comprehensive}. For example, AlphaMix\cite{sun2023mastering} explored a mixture‐of‐experts (MoE) approach specifically designed for generating higher excess returns. From our perspective, the additional alpha stems from a lower epistemic uncertainty, resulting in a left shift in the risk-return plane, pushing $\theta_s$ closer to  PML. While we do not delve further into ensembling in the present work, this connection highlights the inherent and inseparable relationship between risk and return in foundation model trading.

In this section, we developed a Bayesian approach to estimating the Pretrained Market Line, leveraging MC Dropout to disentangle each custom strategy’s risk. The key assumption is that any strategy with a given return can, in principle, be matched by an optimal fine‐tuning that eschews idiosyncratic variance and retains only the portion of risk truly “priced” by the foundation model. By regressing these “priced” risks against observed returns across multiple $\theta_s$, we arrive at an empirical estimate of the foundation Sharpe ratio—the slope of the PML. In the next section, we implement this methodology on real data, illustrating both its strengths and potential pitfalls in practice.

\section{Experiments}

We now present a series of empirical investigations designed to evaluate our CAPM‐inspired framework for foundation‐model trading.  Specifically:

\begin{enumerate}
    \item We first conduct a preliminary study of raw prediction performance of foundation model’s signals at various resolutions, clarifying why we ultimately focus on a 1 s horizon for subsequent experiments.
    \item Building on this, we examine how fine-tuned strategies distribute across the risk–return plane and demonstrate PML estimation for each pretrained model. A windowed PML estimation is performed to better observe model decay over time.
\end{enumerate}

We begin by discussing the data and trading task that serve as the common foundation for each experiment. Unless stated otherwise, identical data sources and methodological choices are used throughout to ensure coherence and comparability across experiments.

\textbf{Data}: We draw upon high‐frequency market data from both U.S. equities and the Binance cryptocurrency exchange, detailed in \autoref{tab:bbo_datasets}:

\begin{itemize}
    \item\textbf{US Equity}: One‐second aggregates of the National Best Bid and Offer (NBBO) for a selection of S\&P 500 and Russell 2000 constituents, collected through the ibkr API\footnote{https://www.interactivebrokers.com/campus/ibkr-api-page/twsapi-doc/\#hist-bid-ask}.
    \item\textbf{Cryptos}: Best Bid and Offer quotes for FDUSD‐quoted pairs on Binance\footnote{ https://developers.binance.com/docs/binance-spot-api-docs/web-socket-streams\# diff-depth-stream}, available at 100 ms intervals.
\end{itemize}

\begin{table}[h]
    \centering
    \caption{Statistics of BBO dataset}
    \label{tab:bbo_datasets}
    \begin{tabular}{lccc}
        \toprule
         & S\&P 500 & Russell 2000 & Crypto \\
        \midrule
        Number of assets & 503 & 1947 & 114 \\
        Start Time                & 2023-01-03 & 2023-01-03 & 2023-01-01 \\
        End Time                  & 2025-01-31 & 2025-01-31 & 2025-01-31 \\
        Resolution                & 1s & 1s & 100ms \\
        \bottomrule
    \end{tabular}
\end{table}

Depending on resolution requirements, we further aggregate or sample these datasets as necessary, enabling in-depth investigations into short-horizon predictive tasks across a range of market settings.

\textbf{Trading Task}: Although our theoretical framework accommodates both single-asset and multivariate (portfolio-level) trading strategies, for the simplicity we restrict our scope to single-asset trading. This choice avoids the added complexity of portfolio aggregation effects, thus allowing a clearer view of each model’s behavior.

\textbf{Models}: We assess a suite of modern, pretrained time series architectures (TimesFM\cite{das2024decoderonly}, Chronos\cite{ansari2024chronos}, Moirai\cite{woo2024unified}, Timer\cite{liu2024timer}, and TTM\cite{ekambaram2024tiny}), alongside conventional baselines (LSTM\cite{nelson2017stock}, DARNN \cite{qin2017dualstage}, MLP\cite{naeini2010stock}, SFM\cite{zhang2017stock}, and GRU\cite{shen2018deep}). All models are either freshly trained or fine-tuned on our collected dataset.

\textbf{Strategy Construction}: Each model outputs a predicted mid-price. (In the case of models generating a patch, we adopt a simple arithmetic aggregation of the patch as the final prediction.) Whenever the predicted mid‐price exceeds the prevailing market quote, we initiate a BUY order; conversely, if the predicted mid‐price falls below the quote, we enter a SELL order. Our backtest assumes orders are filled at the next tick’s best ask—a convention frequently described as a tick-by-tick backtest or immediate execution on next-tick quotes.We then vary hyperparameters (signal thresholds, stop-losses, and take-profit levels) to generate multiple configurations for each model. Each configuration thus occupies a unique position in the risk–return space.

\subsection{Preliminary : Raw Predictive Performance at Multiple Resolutions}

Before delving into risk–return analyses, we first gauge the raw predictive strength of large pretrained models at different time resolutions (from 100ms to 1day), We center this investigation on the notion of model surprise—the difference between the model’s predicted price and the current quote—and measure how strongly this surprise correlates with subsequent market returns. A sustained positive correlation would indicate that the model consistently anticipates price movements, whereas a weaker or transient correlation might imply limited efficacy over shorter trading horizons.

\begin{figure}[h]
	\centering
	\includegraphics[width=0.9\linewidth]{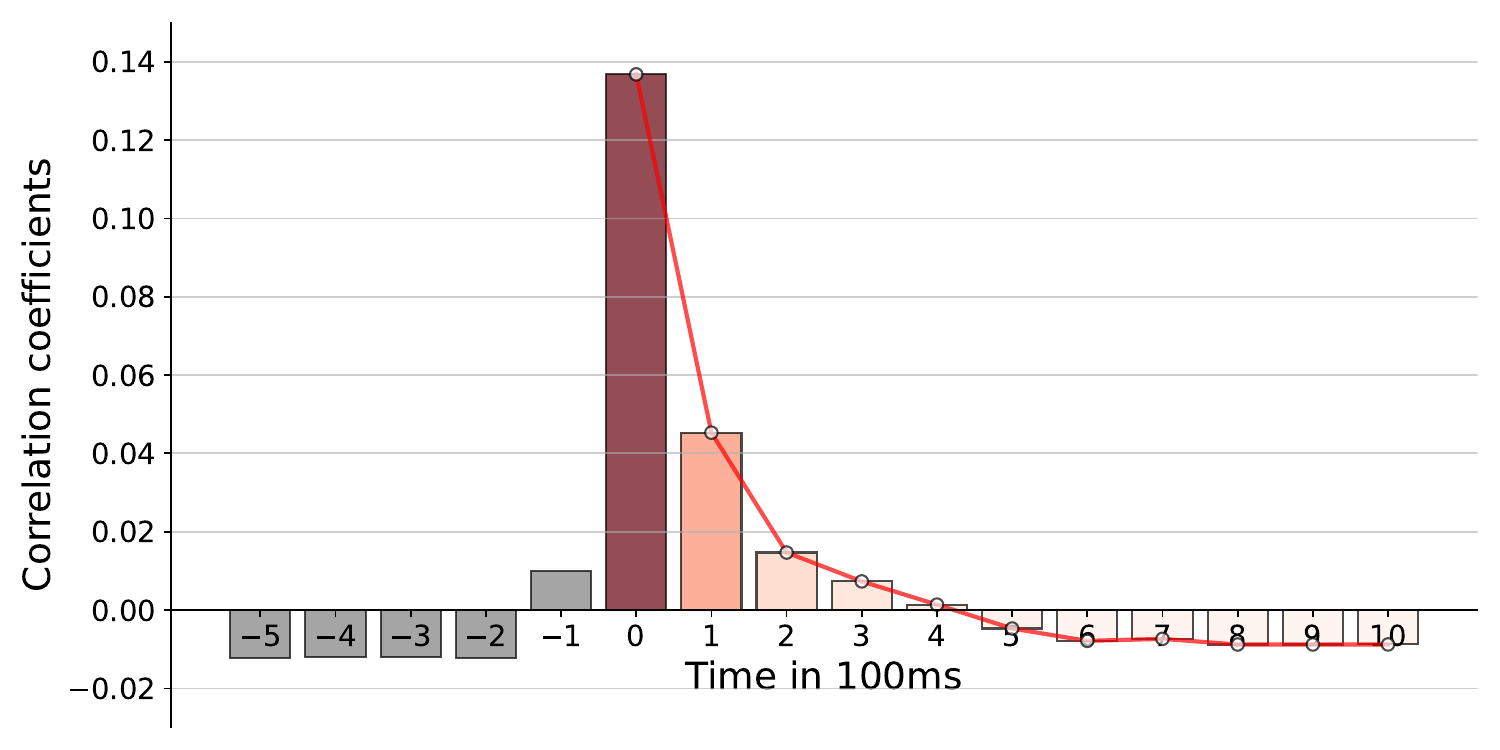} \\
	\includegraphics[width=0.9\linewidth]{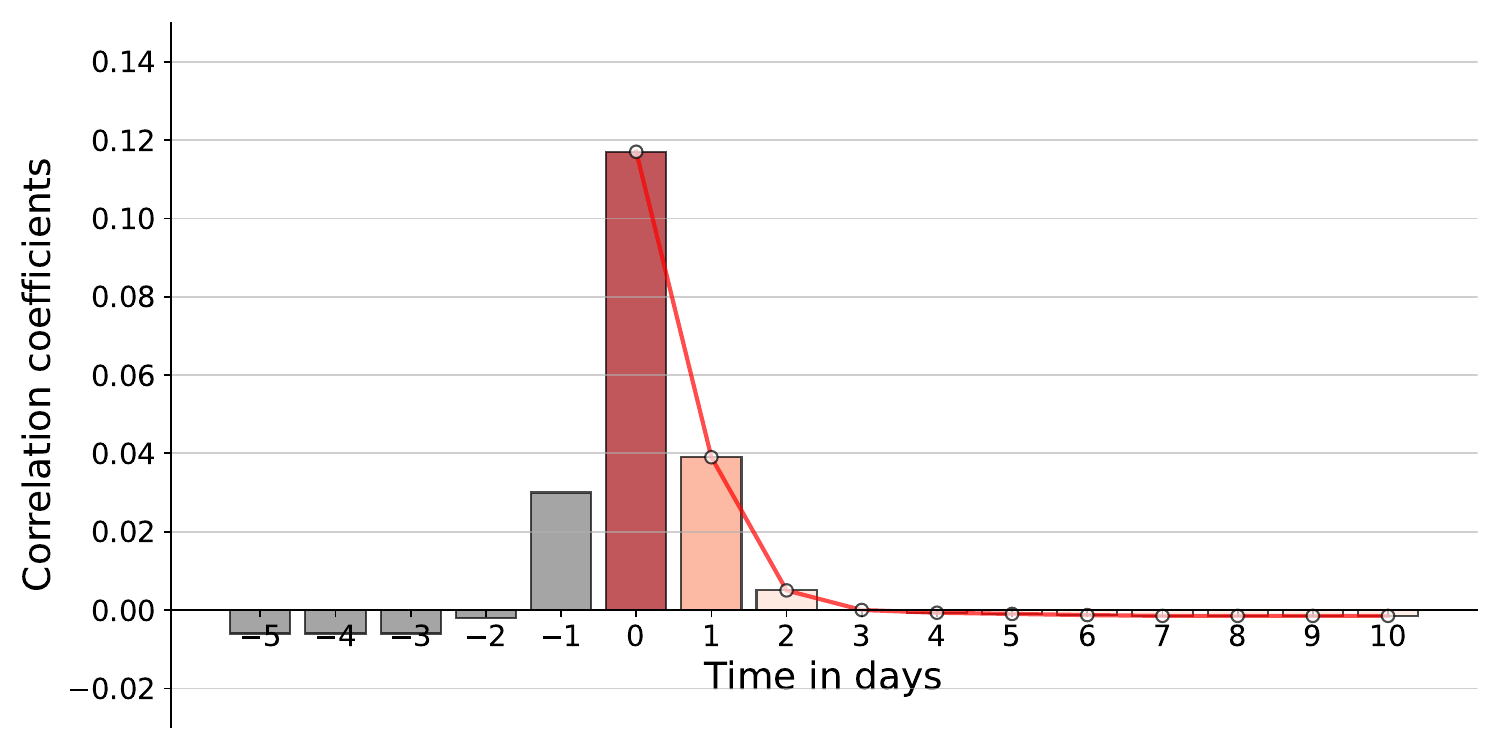}
	\caption{TimesFMv2's model surprise–market return correlation at two prediction resolutions for cryptos.
Top: Correlation coefficients for an ultrafast (100ms) prediction horizon, capturing rapid market reactions. Bottom: Corresponding coefficients for a 1 day horizon, illustrating how the model surprise metric behaves at extended timescales.}
    \Description{Two bar charts with red line overlays show correlation coefficients between the model’s “surprise” and cryptocurrency returns at 100 ms (top) and 1day (bottom) horizons. In the 100 ms chart, correlation peaks around 0.14 near time 0, drops to near zero at about 400 ms, and turns slightly negative thereafter. The 1day chart similarly peaks near 0.13 at time 0, falls to near zero by day 3, and then becomes negative for longer lead–lag times.}
	\label{fig:resolution}
\end{figure}

\begin{figure*}[h]
	\begin{subfigure}{0.49\textwidth}
		\centering
		\includegraphics[width=\linewidth]{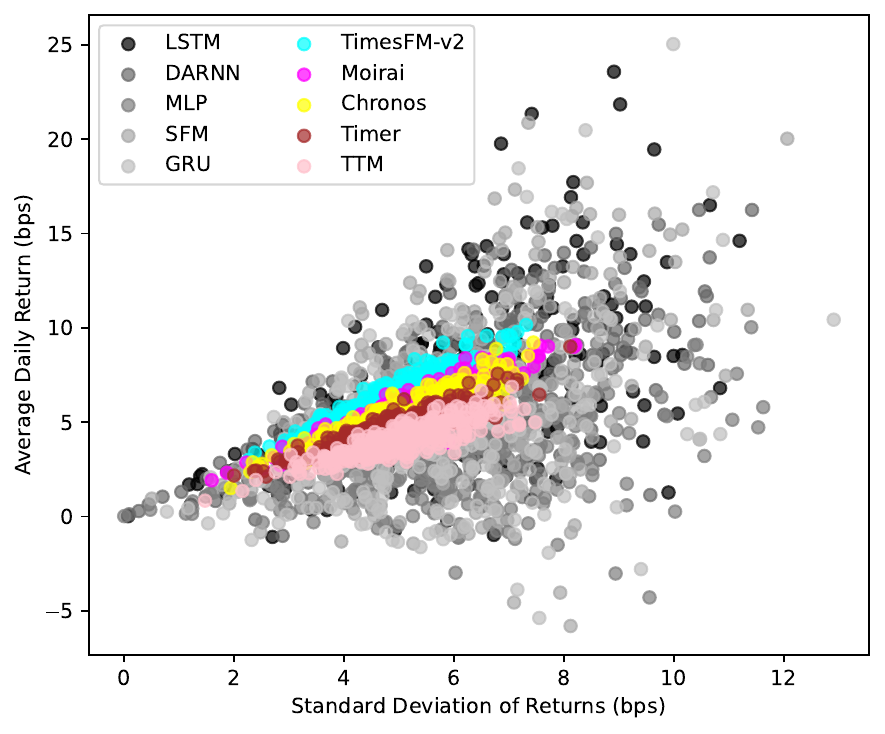}
		\caption{ pretrained strategies (colored) vs baseline (grey)}
		\label{fig:cluster_scatter}
	\end{subfigure}
    \hfill
	\begin{subfigure}{0.49\textwidth}
		\centering
		\includegraphics[width=\linewidth]{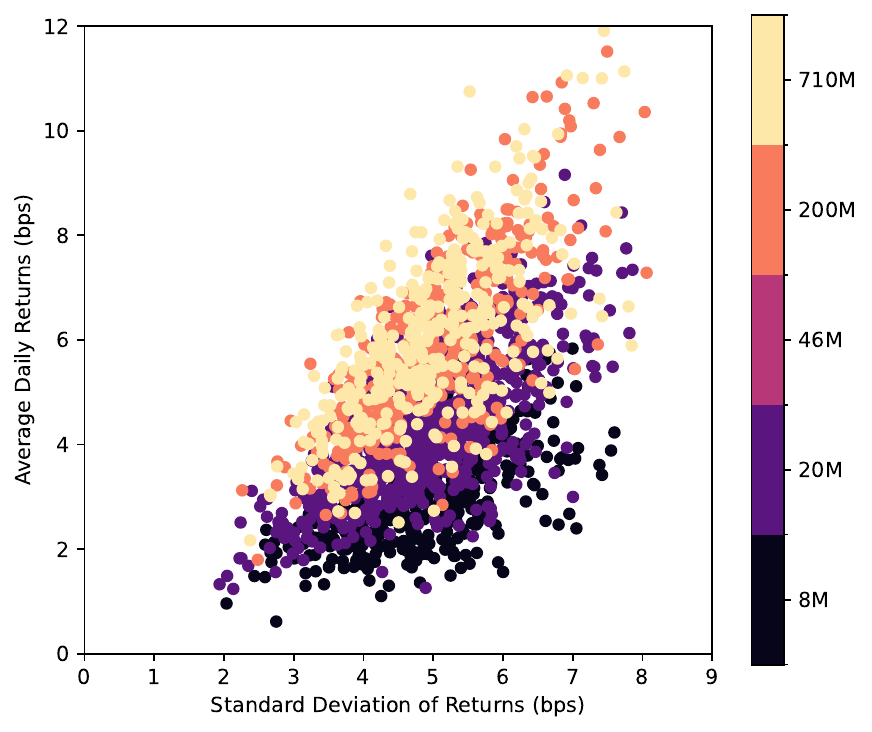}
		\caption{ param scaling of Chronos}
		\label{fig:chronos_scatter}
	\end{subfigure}
    \caption{Daily Risk–Return outcomes (in bps) for all evaluated strategies. (a) Grey markers represent non‐pretrained baseline strategies, and colored markers correspond to foundation‐model–based approaches. (b) A more granular view of Chronos at five parameter scales (8M to 710M), illustrating how model size influences risk–return trade‐offs.}
    \Description{Two scatter plots compare daily risk‐return outcomes (in basis points) across a range of trading strategies. In (a), the x‐axis is the standard deviation of returns (bps) and the y‐axis is the average daily return (bps). Non‐pretrained baseline strategies appear as grey circles (LSTM, DARNN, MLP, SFM, GRU), while pre‐trained foundation‐model‐based approaches are shown in bright colors (TimesFM‐v2 in cyan, Moirai in magenta, Chronos in yellow, Timer in red, and TTM in pink). The data generally exhibit an upward trend, indicating higher returns alongside higher variability. In (b), Chronos is shown at five parameter scales (from 8M to 710M parameters), with color shading from black to yellow denoting increasing model size. This highlights how scaling up Chronos improves its risk‐return trade‐off.}
\end{figure*}

To conduct this analysis, we select TimesFMv2, one of the latest large time series model, and fine-tune it separately for five distinct time resolutions: 100ms, 1s, 5min, 15min, and 1day. We then compute an equal‐weighted average of the correlation between each assets’ model surprise and observed returns. For brevity, \autoref{fig:resolution} showcases only the two extremes—100ms and 1day. We adopt the observational window used in HFT price discovery \cite{naeini2010stock}, capturing correlations from five ticks prior to the future, contemporaneously, and up to five ticks beyond it.

Across all tested resolutions, we observe an initially positive correlation between model surprise and asset returns that progressively diminishes over subsequent ticks. Notably, the correlation often approaches or even falls below zero shortly after predictions are issued, suggesting that any informational advantage is rapidly eroded by better-informed traders. When comparing performance across timescales, the 100ms resolution consistently yields the strongest results—an indication that this high-frequency setting captures the greatest degree of exploitable price predictability. Indeed, we also find exceptionally high correlations in sub-200ms intervals; however, such ultrafast trading windows may be beyond the practical reach of many market participants, as most large pretrained time series models can be efficiently inferenced on GPUs in under a second \cite{ekambaram2024tiny}. Consequently, we adopt 1s as the standard resolution for subsequent experiments, this resolution strikes a balance between practical execution and capturing meaningful intraday signals, plus enabling a consistent comparison between equity and crypto strategies within a unified time scale.

\subsection{Risk–Return Clustering and Foundation Sharpe Ratio}

We next investigate how a broad family of fine‐tuned strategies derived from distinct foundation models distribute in the risk–return space. We also estimate the Pretrained Market Line (PML) slope, which we refer to as the Foundation Sharpe Ratio, using our uncertainty disentanglement approach. 

In this study, we assess five representative pretrained time series architectures—namely TimesFM, Timer, Moirai, Chronos, and TTM—alongside conventional baselines (LSTM, SFM, GRU, ALSTM, and MLP). Each model is either freshly trained or fine-tuned on 1s dataset. To capture a variety of trading behaviors, we then randomize key hyperparameter settings (signal thresholds, stop‐losses, take‐profit levels, etc.), thus generating a diverse spectrum of single‐asset strategies. For each resulting strategy, we record two primary metrics over the backtest window:
\begin{itemize}
    \item Mean daily return $E[r_s]$\footnote{Specifically, a strategy is applied to each asset (SPY and Russell 2000) using an identical bet size, record the resulting returns, and then average these returns to obtain an overall performance measure.}
    \item Volatility of returns $\sigma_s$,  (as a proxy for strategy risk)
\end{itemize}
This setup yields a large collection of $(\sigma_s, E[r_s])$ pairs, laying the foundation for our subsequent analysis of risk–return relationships under different pretrained architectures.

\autoref{fig:cluster_scatter} illustrates the resulting risk–return scatter, where each point corresponds to a unique hyperparameter configuration based on tested bases models ($\theta_\mathrm{TimesFM}$, $\theta_\mathrm{Timer}$, $\theta_\mathrm{Moirai}$, $\theta_\mathrm{Chronos}$, $\theta_\mathrm{TTM}$). Notably, the strategies built upon pretrained foundation models (denoted by colored markers) exhibit a markedly tighter clustering relative to the conventional baselines (represented by grey markers), which are more loosely dispersed across the risk–return plane. This phenomenon suggests that the common informational baseline imposed by the pretrained model tends to homogenize the risk–return characteristics of its fine‐tuned strategies, whereas conventional methods, lacking such a unifying signal, display greater heterogeneity in performance. Moreover, there is non‐trivial overlap among different foundation models themselves, suggesting commonalities in their signals—indeed, multiple large‐scale architectures appear to exploit related market dynamics, which leads to partially correlated outcomes across families.

\autoref{fig:chronos_scatter} zooms in on a single architecture (Chronos) at different model scales (8M to 710M parameters). Larger versions of Chronos systematically shift the cluster left‐upwards—i.e., toward reduced volatility and increased mean return—demonstrating that scaling can materially enhance the quality of the learned signals. However, when combined with the latency‐sensitive nature of these generative signals (Section 4.1), scaling up parameters also increases inference times—particularly significant in high‐frequency contexts—thereby introducing a trade‐off between model capacity and execution speed. Addressing this tension in a systematic manner remains an important open question for future research.

To quantify how much of each strategy’s variance is actually priced by the pretrained signals, we apply Monte Carlo Dropout to separate total risk into epistemic (foundation‐model) and aleatory components. In line with prevailing short‐term U.S. Treasury yields for given period, we fix the risk‐free rate $r_f$ at 5\%, then regress each configuration’s expected daily return on its aleatory volatility. This approach yields Foundation Sharpe ratio, interpreted as the slope of the PML. As shown in \autoref{tab:sr}, the PML slope regression are unstable for tested conventional baselines, suggesting that our notion of “shared risk” is not universally applicable—rather, it becomes most evident in pretrained architectures where common signals genuinely dominate the risk–return profile.

\begin{table}[h]
\centering
\caption{Foundation Sharpe Ratios}
\label{tab:sr}
\begin{tabular}{lccc}
\toprule
& Model         & $\mathrm{SR}_\theta$   & R²    \\ 
\midrule
\textit{Baseline} & LSTM          & $3.44 \pm 0.59$           & 0.30         \\
\textit{Models} & ALSTM         & $3.19 \pm 0.39$           & 0.47         \\
& MLP           & $1.73 \pm 0.37$           & 0.23         \\
& SFM           & $2.85 \pm 0.71$           & 0.17         \\
& GRU           & $2.77 \pm 0.65$           & 0.19         \\
\cmidrule(r){1-4}  
\textit{Foundation} & TimesFM-v2    & $3.87 \pm 0.18$           & 0.86         \\
\textit{Models} & Moirai        & $3.39 \pm 0.16$           & 0.86         \\
& Chronos       & $3.31 \pm 0.17$           & 0.83         \\
& Timer         & $2.77 \pm 0.16$           & 0.80         \\
& TTM           & $2.41 \pm 0.17$           & 0.72         \\
\bottomrule
\end{tabular}
\end{table}

\subsection{Foundation Model Decay}

Finally, we turn our attention to alpha decay—the gradual erosion of excess returns that many widely-used investment strategies experience over time, particularly once they become public knowledge\cite{grinold2000active, penasse2022understanding}. It remains an open question whether foundation-model-based approaches are similarly vulnerable.

If we posit that the sharpe ratio estimated from our PML framework reflects the model’s theoretical limit under optimal fine-tuning, then tracking how this ratio evolves can shed light on how swiftly a foundation model’s signals deteriorate once they become assimilated by the market or outperformed by newer adaptations.

To investigate this question, we re‐fine‐tune a single TimesFM‐v1 model weekly, each time using the most recent 12 months of data. We then backtest the updated strategy at 1s resolution and apply rolling‐window PML estimation to obtain both the observed Sharpe ratio, $\mathrm{SR}_{\theta_s}$, and the theoretical upper bound, $\mathrm{SR}_{\theta}$. In \autoref{fig:decay} below, we also highlight the official publication dates of TimesFM‐v1 and its successor, TimesFM‐v2, to examine whether broader market awareness of these foundation‐model techniques materially affects strategy performance.

\begin{figure}[h]
  \centering
  \includegraphics[width=\linewidth]{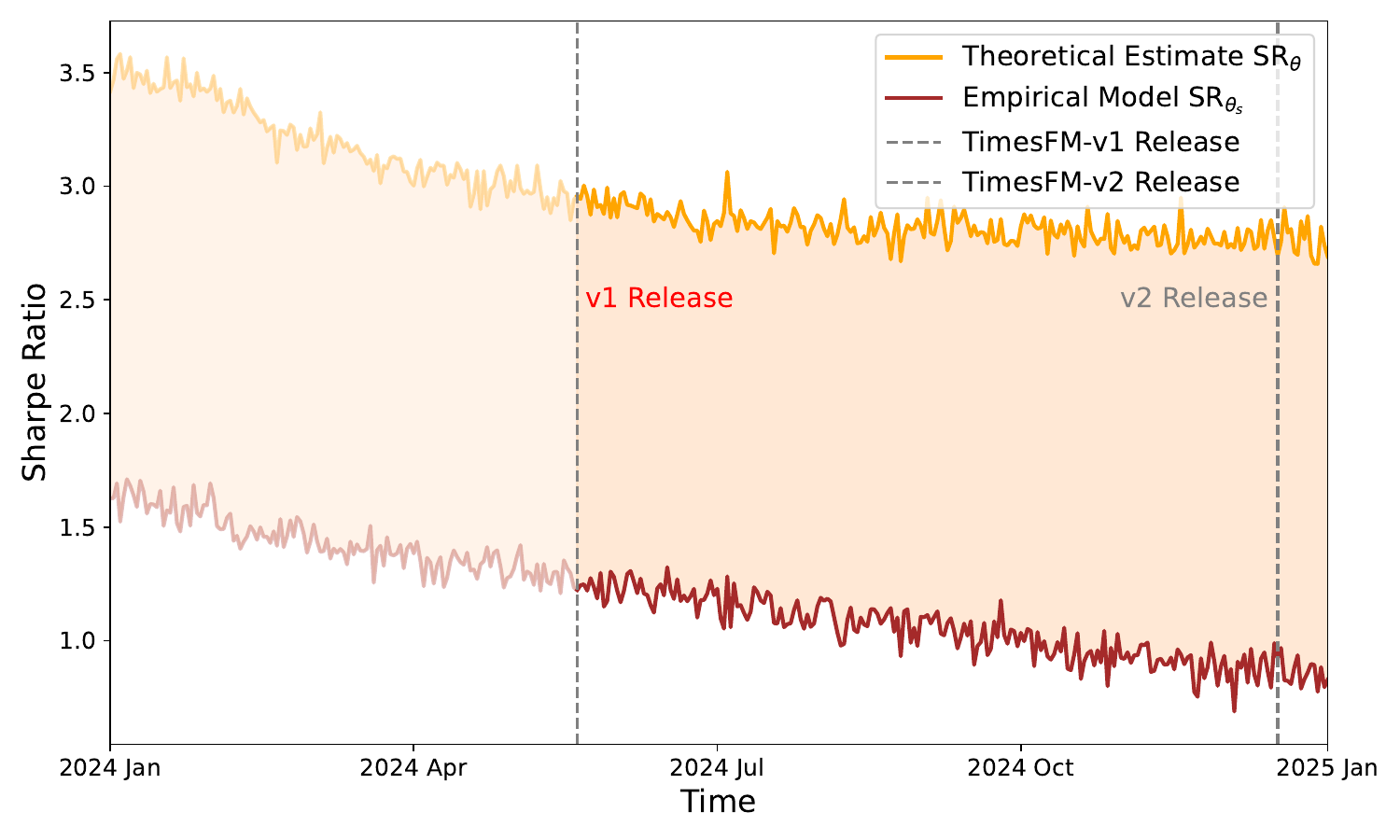}
  \caption{Rolling‐window Sharpe ratios for a TimesFM‐v1‐based strategy, re‐fine‐tuned weekly, from early 2023 to early 2025. The maroon curve ( $\mathrm{SR}{\theta_s}$ ) represents the observed performance, while the gold curve ( $\mathrm{SR}_{\theta}$ ) reflects the theoretical upper bound from PML estimation. Vertical dashed lines denote the public release dates of TimesFM‐v1 and TimesFM‐v2.}
  \label{fig:decay}
  \Description{A two‐line chart shows rolling‐window Sharpe ratios from early 2023 to early 2025. A gold line (theoretical upper bound) and a maroon line (observed performance) both trend downward, with dashed vertical lines marking the releases of TimesFM‐v1 and TimesFM‐v2. The gap between the gold and maroon lines gradually widens over time, illustrating alpha decay. Neither the v1 release in mid‐2024 nor the v2 release in late 2024 produces an abrupt shift in either curve. Instead, the overall decline continues steadily, consistent with the notion of a gradual alpha decay as the market assimilates these strategies.}
\end{figure}

From early 2023 to early 2025, both $\mathrm{SR}_{\theta}$ and $\mathrm{SR}_{\theta_s}$ show a steady decline, hinting that alpha potential (even under optimal tuning) erodes as the market assimilates foundation‐model‐derived signals. This assimilation is further exacerbated by competitors employing more advanced fine‐tuning and execution methods, causing the gap $\mathrm{SR}_{\theta} - \mathrm{SR}_{\theta_s}$ to widen. Interestingly, neither the publication of TimesFM‐v1 nor TimesFM‐v2 immediately produces an abrupt change; the overall downward trend remains consistent, underscoring a gradual but persistent decay in alpha once a foundation model’s edge becomes widely recognized.

Hence, while the ideal returns of foundation‐based trading diminish in tandem with increased market adoption, practical implementations can erode even faster in real‐world conditions. These findings underscore the dynamic nature of alpha generation: any early advantage secured by cutting‐edge foundation models tends to dissipate rapidly as the market refines and extends these strategies.

\section{Limitations and Future Directions}

Below we highlight three key areas where our current work can be further expanded:

\textbf{Multivariate Forecasting and Portfolio‐Level Strategies.} Although our framework theoretically accommodates multi‐asset settings, our experiments focus on single‐asset strategies to maintain clarity. Extending to cross‐asset interactions and portfolio‐level decision‐making remains to be tested and could validate whether our CAPM‐inspired decomposition truly holds at scale. Such an expansion would also illuminate how systematic and idiosyncratic risks propagate across a diverse basket of instruments.

\textbf{Latency–Scaling Trade‐Offs and Lightweight Architectures.} As shown in \autoref{fig:chronos_scatter} and \autoref{fig:resolution}, while larger foundation models generally exhibit stronger predictive signals, they also increase inference latency, potentially negating gains in fast‐moving markets. Future research might investigate quantization, distillation, or pruning strategies to retain model quality in sub‐second environments. Balancing parameter scale against speed and execution costs is a pressing challenge with direct practical implications.

\textbf{Deeper Insights into Shared Risk Factors.} We treated each family of foundation models independently, yet the overlap in risk–return clusters suggests broader shared factors. Investigating cross‐model correlations could reveal more universal risk drivers, shaping an enhanced perspective of how foundation‐model‐derived signals collectively reshape market dynamics.

\section{Related Work}
\subsection{Financial Risk Management}
Financial risk management broadly involves safeguarding an organization’s or investment’s assets from losses arising out of uncertainties such as interest rate fluctuations, market volatility, credit risks, and operational failures \cite{aven2016risk}. Its primary goal is to anticipate these potential threats and mitigate them proactively, ensuring that long-term financial objectives remain attainable.

Value-at-Risk (VaR), first introduced by J.P. Morgan, has long been the de facto standard for financial risk management\cite{simons1996value} despite well-known limitations such as its non-subadditivity and the assumption of normality \cite{Derman1996, berkowitz2011evaluating}. Various advances build on VaR to address these shortcomings, including Modified VaR \cite{favre2002mean} and Expected Shortfall \cite{acerbi2002expected}. Meanwhile, a different lineage of work adopts Bayesian modeling approaches \cite{borison2010manage, lynch2007introduction}, asserting that subjective priors can yield superior insight into tail events. Hybrid frameworks integrating Bayesian inference with VaR \cite{hoogerheide2010bayesian, aussenegg2006uncertainty} further expand the methodological toolkit by updating VaR estimates based on posterior distributions rather than point estimates.

Beyond specific risk modeling approaches, model risk looms large because risk cannot be directly measured but must be statistically estimated, giving rise to specification and estimation errors \cite{Derman1996, morini2011understanding}. Disagreements between candidate models tend to magnify during market distress, frustrating confidence in any single risk reading \cite{danielsson2016model}. In practice, therefore, risk managers often combine multiple risk measures, conduct regular stress testing, and follow ongoing validation procedures as critical lines of defense against potential model failures \cite{darbyshire2012hedge, USFedOCC2011}.

\subsection{Uncertainty Modeling for ML}

The idea of representing uncertainty traces back to the roots of Bayesian inference, where the primary goal is to model unknowns probabilistically rather than rely on point estimates. In the context of neural networks, pioneers like MacKay \cite{mackay1992practical} and Neal \cite{neal2012bayesian} advocated using Bayesian principles to capture the posterior over model parameters, producing uncertainty estimates alongside predictions. As deep learning grew, these ideas transitioned into practical methods to quantify uncertainty in large‐scale models. Notable distributional approaches include MC Dropout \cite{gal2016dropout}, which uses stochastic dropout at test time to approximate a Bayesian ensemble, and Deep Ensembles \cite{lakshminarayanan2017simple}, where multiple independently trained networks capture model variability. More recent refinements—such as SNGP \cite{liu2020simple}, SWAG \cite{maddox2019simple}, Laplace approximations \cite{daxberger2021laplace}, and Evidential Deep Learning \cite{sensoy2018evidential}—explicitly model a second‐order predictive distribution over class probabilities and then aggregate those into scalar uncertainties. Alongside these, deterministic methods(DUQ\cite{van2020uncertainty}, DDU\cite{mukhoti2023deep}, Tempreture Scaling\cite{guo2017calibration}) have emerged that directly predict a single scalar uncertainty without modeling a full posterior distribution.

While classical Bayesian theory primarily yielded one overall measure of uncertainty, proposed technique like information‐theoretical \cite{depeweg2018decomposition} and Bregman decompositions \cite{wimmer2023quantifying} trys to disentangling it into finer sources: aleatoric (irreducible data‐inherent noise) and epistemic (model uncertainty due to limited or imperfect training data). However, recent benchmark \cite{mucsanyi2024benchmarking} suggest that these disentanglements often remain highly correlated, highlighting the need for specialized uncertainties tailored to specific tasks.

Thus, in our paper, the CAPM-style risk analysis provides a financial treatment of disentangling foundation models' uncertainty.

\section{Conclusion}

In conclusion, this work adapted CAPM principles to foundation‐model‐based trading by aligning the model’s epistemic and idiosyncratic uncertainties with the familiar decomposition of market‐wide versus asset‐specific exposures. By leveraging Monte Carlo dropout on large time‐series models for trading U.S. equities and cryptocurrencies, we disentangled the truly priced risk inherent in the pretrained architecture from the unpriced variance introduced by suboptimal fine‐tuning. Our findings show that this shared model‐driven uncertainty closely mirrors CAPM’s concept of systemic risk, providing clearer insights into trading strategies’ risk–return profiles and highlighting how alpha can deteriorate as more market participants adopt these powerful pretrained signals.

\bibliographystyle{ACM-Reference-Format}
\bibliography{sample-base}

\end{document}